\documentclass[useAMS,usenatbib]{mn2e}
\usepackage{graphicx}

\title[Short-Hard GRB 070724A]{Limits on Radioactive-Powered Emission Associated With a Short-Hard GRB 070724A in a Star-Forming Galaxy}

\author[Daniel Kocevski et al.]{Daniel Kocevski$^{1}$, Christina C Th\"one$^{2}$, Enrico Ramirez-Ruiz$^{3}$, Joshua S. Bloom$^{4}$ \newauthor Jonathan Granot$^{5}$, Nathaniel R. Butler$^{4}$, Daniel A. Perley$^{4}$, Maryam Modjaz$^{4,}$$^{6}$ \newauthor William H. Lee$^{7}$, Bethany E. Cobb$^{2}$, Andrew J. Levan$^{8}$, Nial Tanvir$^{9}$, Stefano Covino$^{2}$\\\\
$^{1}$Kavli Institute for Particle Astrophysics and Cosmology, Stanford University, 2575 Sand Hill Road M/S 29, Menlo Park, Ca 94025\\
$^{2}$Istituto Nazionale di Astrofisica, Osservatorio Astronomico di Brera,Via E. Bianchi 46, 23807 Merate, Italy\\
$^{3}$Department of Astronomy and Astrophysics, University of California, Santa Cruz, CA 95064, USA\\
$^{4}$Department of Astronomy, University of California, Berkeley, CA 94720-3411\\
$^{5}$Centre for Astrophysics Research, University of Hertfordshire, College Lane, Hatfield, Herts, AL10 9AB, UK\\
$^{6}$Miller Fellow.\\
$^{7}$Instituto de Astronomia, Universidad Nacional Autonoma de Mexico (UNAM), Apdo. Postal 70-264, Cd. Universitaria, Mexico DF 04510. \\
$^{8}$Department of Physics, University of Warwick, Coventry CV4 7AL\\
$^{9}$Department of Physics and Astronomy, University of Leicester, University Road,Leicester, LE1 7RH, UK}

\begin{document}

\pagerange{\pageref{firstpage}--\pageref{lastpage}} \pubyear{2009}
\maketitle
\label{firstpage}

\begin{abstract}
We present results of an extensive observing campaign of the short
duration, hard spectrum gamma-ray burst (GRB) 070724A, aimed at
detecting the radioactively-powered emission that might follow from a
binary merger or collapse involving compact objects.  Our multi-band
observations span the range in time over which this so-called
Li-Paczy\'{n}ski mini-supernova could be active, beginning within 3 hours 
of the GRB trigger, and represent some of the
deepest and most comprehensive searches for such emission.  We find no evidence
for such activity and place limits on the abundances and the lifetimes
of the possible radioactive nuclides that could form in the rapid
decompression of nuclear-density matter.  Furthermore, our limits are
significantly fainter than the peak magnitude of any previously detected broad-lined
Type Ic supernova (SN) associated with other GRBs, effectively ruling out a long
GRB-like SN for with this event. Given the
unambiguous redshift of the host galaxy ($z=0.456$), GRB~070724A
represents one of a small, but growing, number of short-hard GRBs for which firm
physical/restframe quantities currently exist. The host of
GRB~070724A is a moderately star-forming galaxy with an older stellar
population component and a relatively high metallicity of 12+log(O/H)$_{KD02}=9.1$.  We find no significant evidence for large amounts of
extinction along the line of sight that could mask the presence of a
SN explosion and estimate a small probability for chance alignment with the putative
host.  We discuss how our derived constraints fit into the
evolving picture of short-hard GRBs, their potential
progenitors, and the host environments in which they are thought to be
produced.
\end{abstract}


\section{Introduction}
Long duration gamma-ray bursts (GRBs), lasting more than $\sim 2$
seconds \citep{Kouveliotou93} are thought to originate from the
collapse of massive stars (see Woosley \& Bloom 2006).  Short-duration, hard spectrum, gamma-ray bursts (SHBs), with a duration of less than $\sim 2$ seconds have long been assumed to have a different astrophysical origin than
long-duration events \citep{nakar07,lrr07},
namely the coalescence of compact binaries, the most widely discussed
being neutron star binaries (NS-NS) or a neutron star and a black hole
binary (NS-BH) \citep{bp86,Eichler89,bp91,Narayan91}. The discovery of
afterglows associated with SHBs \citep{Gehrels05, Hjorth05a} led to
the inference that they are associated with an older population of
stars \citep[e.g.][]{Bloom06}.  Subsequent follow-up observations
supported the idea that these events would not be accompanied
by supernovae (SNe) \citep{Hjorth05a, Bloom06}.

Nominally, without explosive nucleosynthesis of $^{56}$Ni to form a
supernova, there should be no late-time optical emission after the
afterglow has faded.  However, during a NS-NS or NS-BH merger,
dense material stripped from the star has been predicted to form large tidal
tails \citep{rosswog07}. Depending on the details of the encounter and the neutron star
equation of state, a fraction of this can be dynamically ejected from
the system. The subsequent decompression of this material could synthesize radioactive
elements through the r--process \citep{rosswog99,Frei99}, whose
radioactive decay could power an optical transient
\citep{LiPaczynski98}. The fraction of material that remains bound
will eventually return to the vicinity of the compact object, with
possible interesting consequences for late-time emission
\citep{rosswog07,lrr07}.


For a number of SHBs, late times limits on additional light arising from Type Ib/c SNe, which have been associated with some long-GRBs, have been obtained: 050509B \citep{Hjorth05b, Bloom06}, 050709 \citep{Hjorth05b, Fox05}, 050724
\citep{Malesani07}, 051221A \citep{Soderberg06}, 050813
\citep{Ferrero07}, 060502B, where
the limits range from 1.5 to over 6 magnitudes fainter than GRB-SN
1998bw. Limits on a mini-SN like scenario at early times ($\sim1$ day)
have so far only been derived for GRB~050509B \citep{Hjorth05b,
  Bloom06}, setting an upper limit of $\leq 10^{-5}$ on the
fraction of rest mass energy that goes into the radioactive decay.  Though
an optical Li-Paczy\'{n}ski mini-supernova (LP-SN) \citep{LiPaczynski98} like bump was seen in GRB 080503 (Perley et al. 2008),
concurrent X-ray emissions suggested a synchrotron powered-afterglow,
rather than a radioactive powered event, better accounted for the
physical origin.  Furthermore, no redshift was available for that
event, so the energetics of the additional emission contributing to the 1 day bump are unconstrained.

Here, we report on new LP-SN limits of a SHB.  GRB~070724A was detected by the Swift satellite on July 24,
2007, 10:53:50 UT and consisted of a single peak with a
duration of T$_{90}$=0.4$\pm$0.04 s \citep{Ziaeepour07,
  Parsons07}. The XRT instrument onboard Swift revealed a counterpart
77.8 arcsec from the center of the BAT position \citep{Parsons07,Page07}. No optical afterglow was detected by the UVOT instrument
\citep{DePasquale07} nor from ground based observations
\citep{Cenko07, Covino07}. A nearby DSS source, 0.5 arcsec from the
center of the XRT position was quickly proposed as a possible host
\citep{Bloom07a, Bloom07b} with a redshift of $z=0.457$
\citep{Cucchiara07} and found not to be varying \citep{Covino07,
  Cucchiara07}. Four radio sources were also found inside the BAT
error circle, although none showed any sign of variability
\citep{Chandra07}.  The field containing the host galaxy was visible
near the end of the night over most of North America, allowing for
comprehensive follow-up observations by a variety of telescopes over
the course of several weeks.

The paper is organized as follows. In $\S 2$ we present a summary of
our observations.  The results of the mini-SN modeling are presented in
$\S 3$, while the properties of the suggested host galaxy and
surrounding galaxy population is discussed in $\S 4$. Finally, the
implications of our results are presented in $\S 5$. Throughout the
paper we assume $H_0 = 71$~km~s$^{-1}$~Mpc$^{-1}$ and a $\Lambda$CDM
cosmology with $\Omega_{\rm m} = 0.27$, $\Omega_\Lambda = 0.73$.

\section{Observations $\&$ Analysis}
\subsection{Swift BAT and XRT}

The detection of GRB~070724A prompted an automated slew of the Swift
spacecraft followed by XRT observations of the field beginning at $T +
72.1$ sec and which continued for roughly $\sim10^{6}$ sec, at which
point the source faded below the detector's sensitivity threshold.  We
obtained the Swift~BAT and XRT data on GRB~070724A from the
Swift~Archive\footnote{\tt ftp://legacy.gsfc.nasa.gov/swift/data}. The
data were processed with version 0.11.4 of the {\tt xrtpipeline}
reduction script from the HEAsoft~6.3.1\footnote{\tt
  http://heasarc.gsfc.nasa.gov/docs/software/lheasoft/} software
release, where we have employed the latest (2007-07-09) calibration
files at time of writing.  The reduction of XRT data from cleaned event lists output by
{\tt xrtpipeline} to science ready light curves and spectra is
described in detail in \citet{ButlerKocevski07a}.

Our best estimate of the position of the XRT detection is $\alpha =
01^{\rm h}51^{\rm m}13''.99, \delta=-18^{\circ}35'39''.1$ with an
error of $\sim 2$ arcsec.  The fluence in the 15-150 keV BAT energy
band is $3.0~\pm~0.7 \times 10^{-8}$ erg cm$^{-2}$.  This is a extremely low value when compared to other Swift detected GRBs, being among the faintest $~2\%$ of Swift bursts and the faintest $20\%$ of Swift  detected short bursts.  Given the redshift $z=0.457$, reported by \citet{Cucchiara07}, the
isotropic-equivalent energy released in gamma-rays is estimated at
$E_{\rm iso} = 1.55 \times 10^{49}$ ergs.

\begin{figure}
\includegraphics[width=89mm]{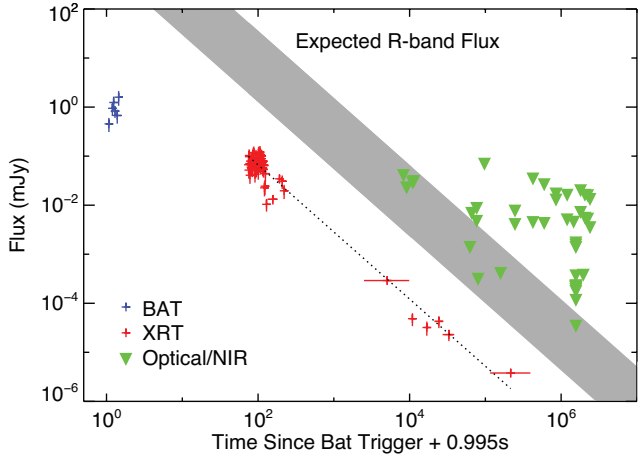}
\caption{Gamma-ray (blue) and X-ray (red) observations of 070724A
  along with our optical and NIR (green) upper limits.  The grey
  region represents the expected $R$-band flux from a standard forward
  shock model given the afterglow's X-ray properties.  Our optical observations rule out a bright optical afterglow for this event}
\label{Fig:limits}
\end{figure}

The XRT light curve (Figure \ref {Fig:limits}) is consistent with an
unbroken power law with a decay index ($F_{\nu} \propto t^{-\alpha}$)
of $\alpha_{X} = 1.37 \pm 0.03$ over the entire span of the Swift
observations.  The windowed timing mode spectrum from $t=74$ sec to
4.1 ksec after the burst is well fit by a power law with photon index (d$N$/d$E$ $\propto E^{-\Gamma}$) of $\Gamma=1.5$ or correspondingly an energy index ($F_{\nu} \propto \nu^{-\beta}$), of $\beta_{X} =0.5$ and photoelectric
absorption yielding $N_H= 2.4 \pm 0.9 \times 10^{22}$ cm$^{-2}$,
significantly greater than Galactic ($N_{\rm H, Galactic}= 1.2 \times
10^{20}$ cm$^{-2}$) \citep{Dickey90}.

\begin{table}
 \caption{Photometric observations of GRB~070724A}
 \label{photometry}
 \begin{tabular}{llllll}
 \hline  
 &  &  & Exp & Limiting & Flux \\
Day & Instrument & Filter & (s) & Mag & ($\mu$Jy) \\
\hline 
   0.0961 & UKIRT/UFTI & $K$ & 540	 & 18.05 & 40.17 \\ 
    0.1057 & UKIRT/UFTI & $J$ & 405 & 19.62 & 22.54 \\ 
    0.1270 & UKIRT/UFTI & $H$ & 270 & 18.88 & 28.65 \\ 
    0.1316 & UKIRT/UFTI & $K$ & 540 & 18.35 & 30.36 \\ 
    0.7254 & NOT/StanCam & $i$ & 900 & 23.47 & 1.38 \\ 
    0.7384 & NOT/StanCam & $R$ & 1800 & 23.72 & 1.01 \\ 
    0.7640 & NOT/StanCam & $B$ & 900 & 21.95 & 6.82 \\ 
    0.8896 & CT1.3m/ANDICAM & $J$ & 60 & 20.67 & 8.59 \\ 
    0.8897 & CT1.3m/ANDICAM & $I$ & 360 & 21.79 & 4.69 \\ 
    0.9271 & VLT/FORS & $i$ & 119 & 24.74 & 0.31 \\ 
    1.1270 & UKIRT/UFTI & $K$ & 1440 & 17.47 & 68.53 \\ 
    2.8488 & CT1.3m/ANDICAM & $J$ & 1800 & 20.80 & 7.59 \\ 
    2.8488 & CT1.3m/ANDICAM & $I$ & 2160 & 21.95 & 4.04 \\ 
    4.9158 & CT1.3m/ANDICAM  & $J$ & 1800 & 19.16 & 34.27 \\ 
    4.9158 & CT1.3m/ANDICAM  & $I$ & 2160 & 21.85 & 4.43 \\ 
    6.9235 & CT1.3m/ANDICAM  & $J$ & 1800 & 19.46 & 26.13 \\ 
    6.9235 & CT1.3m/ANDICAM  & $I$ & 2160 & 21.90 & 4.24 \\ 
    9.9057 & CT1.3m/ANDICAM  & $J$ & 1800 & 19.93 & 16.94 \\ 
    9.9057 & CT1.3m/ANDICAM  & $I$ & 2160 & 20.70 & 12.71 \\ 
   13.9056 & CT1.3m/ANDICAM  & $J$ & 1800 & 19.99 & 15.96 \\ 
   13.9008 & CT1.3m/ANDICAM  & $I$ & 2160 & 21.71 & 5.03 \\ 
   16.9389 & CT1.3m/ANDICAM  & $I$ & 2160 & 21.83 & 4.50 \\ 
   18.1155 & Keck I/LRIS & $R$ & 300 & 25.10 & 0.28 \\ 
   18.1155 & Keck I/LRIS & $g'$ & 30 & 23.32 & 1.70 \\ 
   18.1221 & Keck I/LRIS & $R$ & 10 & 23.39 & 1.37 \\ 
   18.1237 & Keck I/LRIS & $R$ & 10 & 23.26 & 1.54 \\ 
   18.1257 & Keck I/LRIS & $R$ & 600 & 25.39 & 0.22 \\ 
   18.1307 & Keck I/LRIS & $g'$ & 200 & 25.00 & 0.36 \\ 
   18.1341 & Keck I/LRIS & $R$ & 600 & 25.37 & 0.22 \\ 
   18.1342 & Keck I/LRIS & $g'$ & 630 & 25.50 & 0.23 \\ 
   18.1422 & Keck I/LRIS & $g'$ & 630 & 25.70 & 0.19 \\ 
   18.1423 & Keck I/LRIS & $R$ & 600 & 25.44 & 0.21 \\ 
   18.1506 & Keck I/LRIS & $R$ & 600 & 25.44 & 0.21 \\ 
   18.1506 & Keck I/LRIS & $g'$ & 630 & 25.80 & 0.17 \\ 
   18.1416 & Keck I/LRIS & $R$ & 2720 & 27.40 & 0.03$^{a}$ \\ 
   18.1460 & Keck I/LRIS & $g'$ & 2120 & 26.25 & 0.12$^{b}$ \\ 
   20.9249 & CT1.3m/ANDICAM  & $J$ & 1800 & 19.77 & 19.76 \\ 
   20.9250 & CT1.3m/ANDICAM  & $I$ & 2160 & 21.33 & 7.13 \\ 
   22.6958 & NOT/StanCam & $R$ & 3600 & 24.80 & 0.37 \\ 
   23.8962 & CT1.3m/ANDICAM  & $J$ & 1800 & 20.00 & 15.95 \\ 
   23.8962 & CT1.3m/ANDICAM  & $I$ & 2160 & 21.63 & 5.43 \\ 
   25.9324 & CT1.3m/ANDICAM  & $J$ & 1800 & 20.02 & 15.65 \\ 
   25.9324 & CT1.3m/ANDICAM  & $I$ & 2160 & 21.67 & 5.20 \\ 
   27.8593 & CT1.3m/ANDICAM  & $J$ & 1800 & 20.21 & 13.10 \\ 
   27.8593 & CT1.3m/ANDICAM  & $I$ & 2160 & 22.11 & 3.48 \\
\hline 
\end{tabular} 
\\We assume a $T_{0}$ of 10:53:50 on 07-24-2007 \\
$^{a}$ Coaddtion of all $R$-band Keck observations \\
$^{b}$ Coaddtion of all $g'$-band Keck observations
\end{table}

We note that there are significant and rapid variations at early times in the X-ray
hardness, which suggest that the excess column is a spurious result of
the simple model assumption and that this emission is not due to the
external shock afterglow (cf$.$ \citealt{ButlerKocevski07a}).  Indeed,
there are no hardness variations after $t\approx 300$ sec, and the
column density for the photon counting (PC) mode spectrum at these
times is consistent with the Galactic value.  The hardness is
consistent with the $\Gamma=2.0$ commonly observed in other XRT detected GRBs, both long and short. \citep{ButlerKocevski07b}.

Assuming $N_{H}=1.2 \times 10^x{20}$ cm$^{-2}$ and $\Gamma=2.0$
($\beta_{X} = -1.0$), the conversion from the $0.3-10.0$ keV count
rate to flux in $\mu$Jy at 1 keV is 0.044 $\mu$Jy cps$^{-1}$. The unabsorbed
flux after $t \approx 300$ sec until $t\approx 550 $ ksec is $F_E$ =
(2.6$\pm$0.4) x 10$^{-5}$ $(t/[10^3{\rm ~s}])^{-1.37\pm0.03}$ $\mu$Jy.  We
find that the afterglow flux at $t = 10$ hrs is $F_{X,10} \approx 6.0
\times 10^{-14}$ ergs s$^{-1}$, which at a redshift of $z=0.457$ gives
a luminosity of $L_{X,10} = 4.6 \times 10^{43}$ ergs s$^{-1}$.
Assuming that this flux comes from an adiabatically expanding external
shock in the slow-cooling regime, the expected (unabsorbed) $R$-band
optical flux is a factor $\approx 20-520$ times higher, depending on
the location of the cooling break below the X-ray band
\citep[e.g.][]{Sari98}.

The resulting BAT and XRT detections are plotted in blue and red
respectively in Figure \ref{Fig:limits}.  The expected optical flux
from the forward shock falls within the grey region shown in the plot.
Our optical observations discussed below rule out a bright optical afterglow for this event, although a cooling break located just below the XRT bandpass would allow for optical emission to have gone undetected and still fit within the framework of the standard external shock afterglow model.  The lack of a bright optical afterglow for GRB~070724A is consistent with the trend for low-fluence bursts to have faint afterglows noted by \citet{Gehrels08} and \citet{Nysewander08}.  This makes GRB~070724A  particularly interesting because it allows for the search of SN related optical emission without the need to contend with external shock powered afterglow emission.

\begin{figure*}
\includegraphics[width=7in,angle=0]{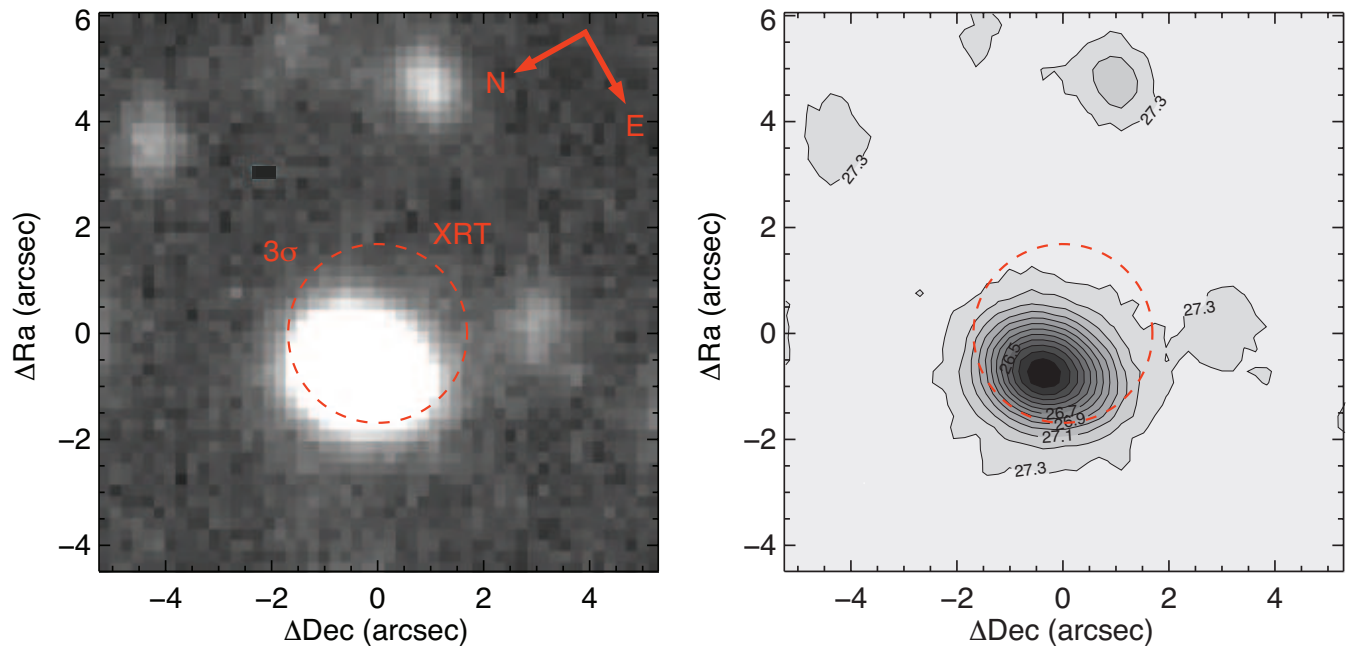}
\caption{(Left) Stacked Keck R-band image (taken 18 days after the event) of the region around the putative host galaxy. The 3 sigma XRT error circle is plotting in red. (Right) Spatially resolved limiting magnitude contour map, dervied after removing an elliptical isophotal model for the host.}
\label{Fig:Lmagmap}
\end{figure*}

\subsection{Optical/NIR}

Our earliest observations of the location of GRB~070724A were taken
with the United Kingdom Infra-Red Telescope (UKIRT) at Mauna Kea,
beginning with a 20 sec exposure in $K$-band at 13:11 UT on July 7th,
roughly 2.3 hours after the trigger.  Several epochs of observations
followed in the $J$, $H$, and $K$ bands over the next 24 hours, ending
approximately 1.1 days after the burst.


Our imaging campaign in the optical passbands began with the use of StanCam at the Nordic
Optical Telescope (NOT) on La Palma/Spain starting at 04:18 UT,
roughly 17.5 hours after trigger.  These observations consisted of
3$\times$300 sec in $B$ and $I$-band and 3$\times$600 sec in
$R$-band. Comparison images of 6$\times$600 sec in $R$ were taken at
03:35 UT on August 16th 2007, 22.7 days after trigger.


We obtained an additional 600 sec $I$-band image with the ESO-VLT
equipped with the FORS2 instrument starting at 09:08 UT (22.1 hours
after trigger), followed by spectroscopic observations, which will be
described below.


Our most extensive observations of the error circle of GRB~070724A
were performed using the ANDICAM instrument operated by the SMARTS Consortium on the 1.3m telescope at
CTIO.  Optical/IR imaging in $J$ and $I$-bands began at 08:14 UT (21.3
hours after trigger) and continued for an additional 8 epochs with the
last observations occurring at 07:31 UT on August 21st (27.8 days post
trigger).  Several dithered images were obtained in each filter, with
total summed exposure times of 30 minutes in $J$ and 36 minutes in $I$.


Finally, we obtained deep optical imaging with the 10m Keck I
Telescope equipped with the LRIS instrument \citep{Oke95} at Mauna
Kea, starting at 13:40 UT on August 11th.  The co-addition of
successive observations yields a total effective exposure time of 2720
seconds in $R$ and 2120 seconds in $g^{\prime}$-bands, followed by
spectroscopic observations, which will be described below.  All
observations were conducted under photometric conditions and
zeropointing was performed using a standard star field observed at an
similar airmass.


\begin{figure*}
\includegraphics[width=7in,angle=0]{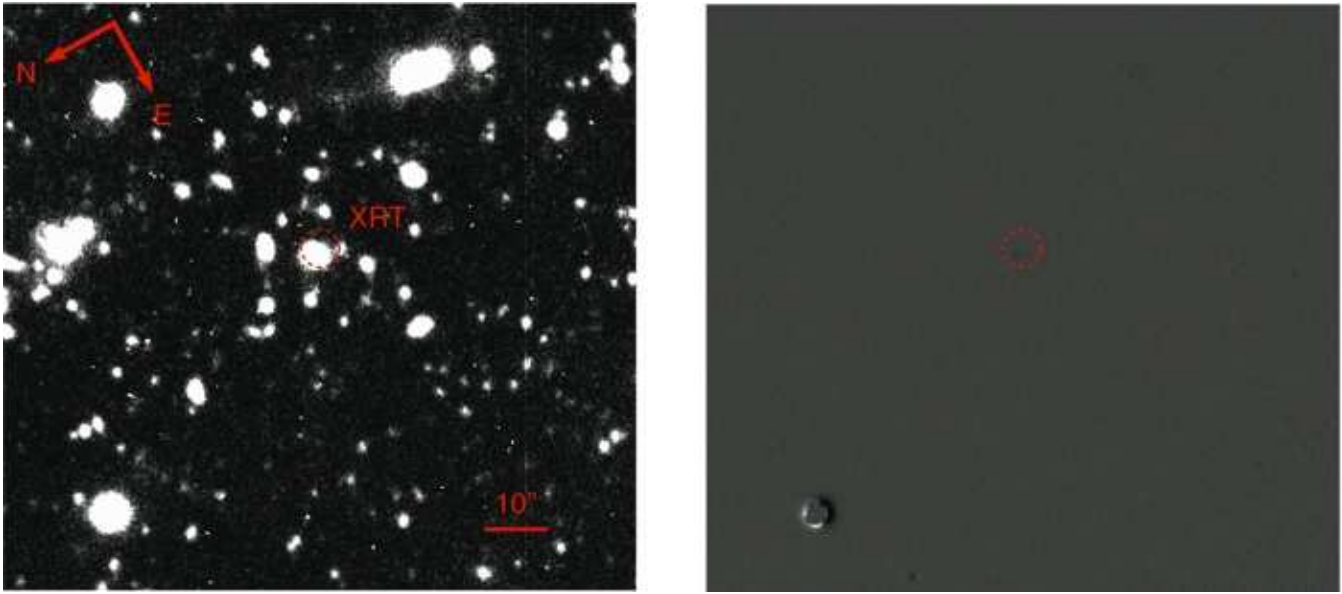}
\caption{(Left) Stacked Keck R-band image (taken 18 days after the event) of the region around the putative host galaxy. The 3 sigma XRT error circle is plotting in red. (Right) The subtraction of the same field field observed one year after the event showing no significant sources at the position of the XRT position.  Similar subtractions were performed for all multi-epoch observations that spanned more than one day.}
\label{Fig:Sub}
\end{figure*}

\subsection{Photometry}

All of our observations were reduced using standard CCD packages in
IRAF.  Once reduced, the co-addition of successive observations was
performed using the {\em Swarp} and {\em Shiftadd} software
packages\footnote{\tt
  http://terapix.iap.fr/rubrique.php?id\_rubrique=49} to produce
weighted sum images of the host field.  Astrometry was performed
relative to USNO-B1 using at least 15 sources in common
between our summed image and the catalog.

A custom pipeline was then used to perform photometry on individual
and coadded frames, using aperture photometry via the {\em Sextractor}
software package \citep{Bertin96}, to estimate instrumental
magnitudes.  An aperture size of $6/5 \times$ seeing (FWHM) was used
for the analysis across all our science data.  Our instrumental
magnitudes were then compared to a standard star field typically taken
on the night of each observations for zeropoint determination of each
reduced frame.  The aperture size used to photometrize the standard
star frames was typically $2-3$ times larger than the aperture used in
the analysis of particularly deep, and hence crowded, images.  For
these images, an aperture correction was applied to account for the
differing aperture radii.

To search for afterglow emission and to properly account for the host
galaxy contribution, we employ a modified version of the public
POIS-IPP package\footnote{\tt
  http://pan-starrs.ifa.hawaii.edu/project/IPP/software/}.

Using the last image in each instrument series as a template of the host, we find
no evidence for residual afterglow emission.  Subtractions employed on images spanning much smaller time scales and
taken with other facilities likewise show no signs of variability
between observations.  A subtraction between the coaddition of all $R$-band Keck observations taken 18 days after trigger and a 30 minute exposure taken 1 year after the event can be seen in Figure \ref{Fig:Sub}.

We estimate the limiting magnitudes on these observations by placing
1000 blank apertures at random positions on the images.  We take the
standard deviation of the resulting photometry distribution to
represent the counts associated with the sky noise of that image and
assume that the faintest observable object is 3 times this value for a
$3\sigma$ limiting magnitude.  The limiting magnitudes as a function
of position on the host galaxy were calculated by taking
this theoretical limiting magnitude due to the sky and adding in
quadrature the error associated with the counts from the galaxy at a
given pixel, which were assumed to be Poissonian and therefore
$\propto \sqrt{N}$.  The resulting spatially resolved limiting
magnitudes are shown as a contour plot in Fig$.$ \ref{Fig:Lmagmap}. We
check these estimates by placing fake stars with a magnitude near our
estimated limiting magnitudes in a subset of our frames and then test
whether they would be detectable at varying thresholds above
background.

The summary of the optical/NIR observations and their associated upper
limits can be found in Table~\ref{photometry}. Figure
\ref{Fig:limits} shows the upper limits at different epochs along with
the BAT and XRT detections.  Our early $R$-band
observations taken with the NOT at $\sim0.73$ days effectively
rule out any emission originating from the forward shock in the
context of the canonical fireball model.

\subsection{Spectroscopy}\label{}
We obtained a single 900 sec spectrum (henceforth called ``slit 1'')
of the host galaxy \citep{Bloom07a} of GRB~070724A with LRIS
(low-resolution imager and spectrograph) at the Keck I telescope,
Mauna Kea, Hawaii, on August 11, 2007. The spectrum covers the
wavelength range between 3500 and 9400 \AA. A slitwidth of 1\farcs0
and grism 600/4000 were used in the blue, providing a resolution
of 4.0 \AA{} FWHM.  In the red, a grism 400/8500 was used providing a resolution of $\sim$ 6.5 \AA{} FWHM. The spectra were extracted and wavelength
calibrated using standard tools in IRAF. All wavelengths given are in
air. Flux calibration was done using the spectrophotometric standard
star BD+174708. The longslit covered both the host galaxy and
a neighboring galaxy to the south-west of the host galaxy at similar
redshift (henceforth called G3).  Note that the narrow slit of
1\farcs0 does not contain the entire flux for both galaxies, which
affects later analyses derived from the flux calibrated spectra.

Furthermore, we obtained spectra with LRIS at three other slit
positions on October 10 and 11, 2007, with the same setting and
calibrated with the standard star Feige 110. The three slits (``slit
3, 4 and 5'') cover a number of galaxies of which only three show
strong emission or absorption lines allowing for redshift determinations.

We also obtained a 600 sec spectra with FORS 1 (called ``slit 2'') at
the VLT on Cerro Paranal, Chile, on July 25, 2007, starting 09:52 UT 
using grism 300V and a 1\farcs0 slit, which has a resolution of 11
\AA{} FWHM. Due to heavy fringing, the emission lines of the host are
not detected with high significance. The slit was positioned in the
north-south direction and therefore covered a slightly different part
of the host galaxy. The other objects in the slit do not show any obvious
emission lines. Reduction and calibrations were also performed with
IRAF standard tasks and flux calibration performed using the
spectrophotometric standard LTT9239. In contrast to the LRIS spectra,
wavelengths for the FORS spectrum are in vacuum.

The position of the slits and the field around the host of GRB~070724A can be seen in Figure \ref{hostpic}.

\section{The host galaxy and its neighbors}
The XRT position of GRB~070724A, with an error radius of 1.6 arcsec, is
offset by roughly 0.8 arcsec from the center of a $z=0.456$, 19.55 $R$-band mag, galaxy that we identify as the host galaxy.  Following the formalism for small offsets (Appendix B in Bloom et al. 2002), we estimate that the chance alignment of a burst given this redshift, offset, and host magnitude is exceedingly small, roughly $P = 0.002$.  At this redshift, the projected offset would transform to a physical distance of 5 kpc from the center of the galaxy although this value is largely unconstrained due to the relative size of the projected offset to the XRT error radius.  Offsets of order of a few kpc have been seen previously in some short bursts, though much larger offsets have frequently been reported \citep{Bloom07c, Troja08}. As short GRBs are assumed to originate from the merger of two compact objects, the kick from the SNe explosions could lead to large offsets from their birthplace and their host galaxies, which seems to agree with the observations (e.g. Fryer, Woosley, Hartmann
1999; Bloom, Sigurdsson, \& Pols 1999; Belczynski et al. 2006). No
other bright sources have been found within or near the XRT position, however,
two galaxies at a similar redshift have been discovered at a distance
of several arcsec from the GRB position.  We analyze the properties of these galaxies together with the host galaxy in this section (see also
Sec. \ref{photozs}).

\begin{figure}
\includegraphics[width=\columnwidth, angle=0]{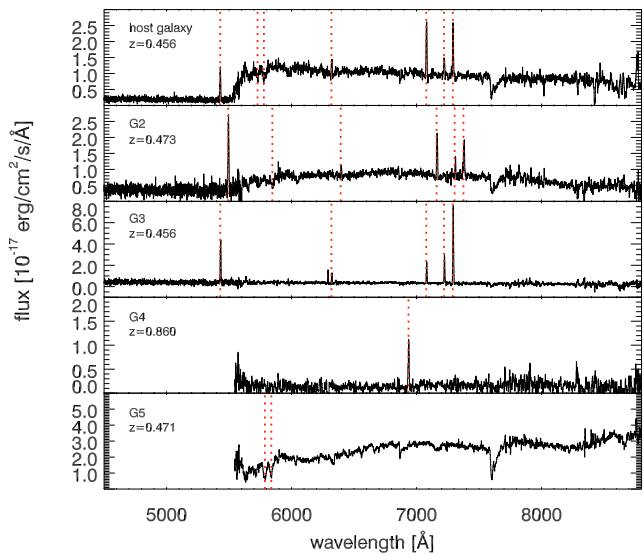}
\caption{Spectra of the host galaxy and some of its neighbours with
  determined redshifts. The lines indicated correspond to the lines
  noted in Table \ref{specplot}.
\label{hostspectra}}
\end{figure}

Figure \ref{hostspectra} shows the spectra of the host as well as all
galaxies where the redshift could be determined, named G2, G3, G4 and
G5 in Figure \ref{hostpic}.

\subsection{Properties of the Host}\label{extinction}

\begin{figure}
\includegraphics[width=\columnwidth, angle=0]{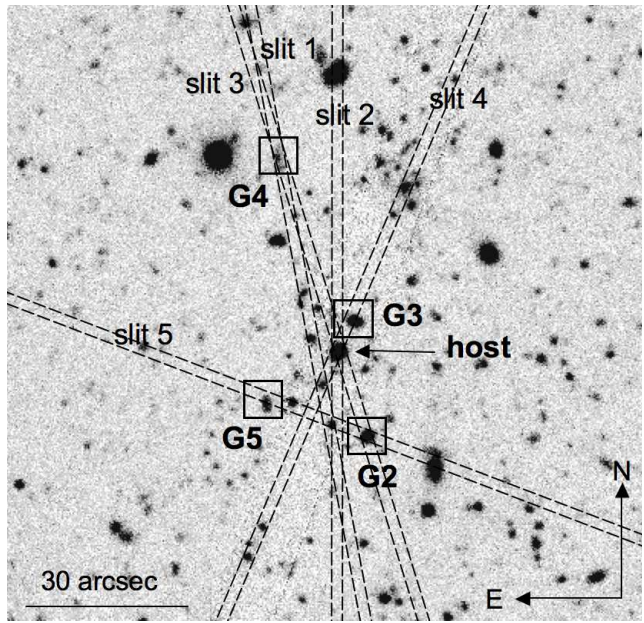}
\caption{Field around the host galaxy of GRB~070724A and positions of
  the slits. The galaxies in the field with measured redshifts are
  indicated according to the notation in Table \ref{hostfield}.
\label{hostpic}}
\end{figure}

The star-forming nature of the host galaxy is suggested by the detection of
several emission lines in the spectra, including [O~{\sc ii}]
$\lambda\lambda$3727,3829, [O~{\sc iii}]~$\lambda\lambda$4959, 5007,
H$\beta$ and H$\gamma$ as well as the Ca H\&K absorption
lines. Unfortunately, the redshift of z$=$0.456 places H$\alpha$ at
$\sim$9600\AA{} outside the wavelength range of our spectra. In the
Table 2, we give the emission line values for the two different slit positions, which
probe slightly different regions, where ``slit 1" indicates the Keck observation and ``slit 2'' indicates the
position of the FORS observation. (in North-South direction).  All other slits shown in Figure \ref{hostpic} were observed with Keck.

\begin{table}
 \caption{Photometric observations of the host galaxy of GRB~070724A}
 \label{photometry:host}
 \begin{tabular}{llllll}
 \hline  
Instrument & Filter &  Mag & Flux ($\mu$Jy) \\
\hline 

NOT/StanCam & $B$ & 21.12 & 9.120 \\
Keck I/LRIS & $g'$ & 21.64 & 8.02 \\
Keck I/LRIS& $R$ & 19.56 & 46.56 \\
CT1.3m/ANDICAM & $I$ & 20.01 & 24.32 \\
UKIRT/UFTI & $J$ & 19.81 & 18.967 \\
UKIRT/UFTI & $H$ & 20.05 & 9.772 \\
UKIRT/UFTI & $K$ & 19.56 & 10.000 \\

\hline 
\end{tabular} 
\end{table}

Using the detected emission lines in slit 1 and slit 2, we can derive a number of properties,
including the extinction, star-formation rate and metallicity of the
star-forming regions. As H$\alpha$ is not available for extinction
measurements using the Balmer line decrement \citep{Osterbrock}, we
use the ratio between H$\gamma$ and H$\beta$, which is 0.47 (for $T_{e}=10^{4}$ K, n$_{e}=100$ cm$^{-3}$) in the absence of any extinction.  Unfortunately, H$\gamma$ is not observed in slit 2, likely due to the lower S/N, so we rely soley on the slit 1 for our extinction estimates.  Using the measured line fluxes for slit 1 shown Table 2 and assuming R$_{\rm V}$ = 3.1 and the \citet{Cardelli89} extinction curve, we obtain a reddening value of E(B$-$V)=1.2 $\pm$ 0.2 mag.  The Galactic reddening along the line of sight towards the host is only
E(B$-$V)=0.013 mag \citep{Schlegel98}. From the detection of the Ca-break and the shape of the spectrum, we conclude that the galaxy has an underlying stellar older population which contributes some absorption in the Balmer lines. This affects H$\gamma$ more than H$\beta$, which, if we were able to correct for it, would lead to a lower value for the extinction. Due to the low resolution of the spectrum, we are not able to determine the strength of the stellar absorption. 

The star-formation rate can be obtained from the H$\alpha$ or the
[O~{\sc ii}] $\lambda\lambda$3727,3729 line flux using the conversion
from \cite{Kennicutt98}. H$\alpha$ is outside of our spectral range
and we therefore use [O~{\sc ii}], which has a more indirect
connection to the ongoing SFR than H$\alpha$. Taking the unextincted values, we then obtain an
absolute SFR of 0.83$\pm$0.03 M$_\odot$/yr from the Keck slit (slit 1) and an
specific SFR scaled with the $g$-band magnitude of 21.75 mag (at
$z=0.456$ equal to restframe $u$-band) of 1.64 M$_\odot$yr$^{-1}$(L/$L$*)$^{-1}$
assuming that an $L$* galaxy has an absolute magnitude of M$_B=-21$
mag \citep{Christensen04}. The extinction corrected fluxes give a rather high absolute SFR or 129$\pm$4 M$_\odot$/yr (the errors from the extinction correction are not propagated) and a SSFR of 253  M$_\odot$yr$^{-1}$(L/$L$*)$^{-1}$. Due to the stellar absorption in the Balmer lines mentioned above, the true SFR lies inbetween the unextincted and the extinction corrected values. 

Using the R$_{23}$ parameter, which takes the ratio of the [O~{\sc
    ii}], [O~{\sc iii}] and H$\beta$ emission line fluxes
(uncorrected for possible extinction in the host), we derive a
metallicity for the host galaxy. The most frequently applied
calibration for GRB hosts is the one taken from \cite{KD02}, which
gives a rather high metallicity of 12+log(O/H)$_{KD02}=9.1$ for this galaxy as compared to the
solar metallicity value of 8.66 \citep{Asplund04}. Using the extinction corrected fluxes leads to the same value for the metallicity. The intrinsic error of this
method for determing the metallicity is around 0.2 dex.

The H$\beta$-EW is dependent on the age of the stellar population and
can be used as an upper limit for the age according to stellar
evolutionary models, e.g. \cite{Leitherer99}. A value of
log(EW$_{H\beta}$)=1.04 or 1.22 for the two slit positions corresponds
to an age of about 10 Myrs for an instantaneous starburst and about
100 Myr for continous star formation depending slightly on the
metallicity. The Balmer break, which is clearly visible in the spectrum
of the host galaxy, furthermore indicates an age of at least 100 Myrs,
indicating an older population of stars.  This is consistent with the photometry for this host and quite interesting given the detection of strong emission features.  It could indicate recent starburst activity, possibly due in part to a merger event with one of its nearby neighbors, revitalizing what was otherwise a old, red galaxy.

\subsection{Is the Host Part of a Cluster?}\label{photozs}
Since the first well-localized SHB (GRB 050509b) was found to be in
a cluster (Bloom et al. 2006), the frequency of SHBs in clusters has
been a subject of study (Berger et al. 2006). To investigate the
potential for diffuse X-ray emission from a cluster of galaxies, we
search for extended sources using {\tt wavdetect} \citep{Freeman02}.
We analyzed the X-ray data on scales of 0.8, 1.1, 1.6, 2.2, 3.1
arcmins and found no strong detections which are not centered on point
sources in the 91.7 ksec exposure XRT PC mode image (Detections
centered on point sources are considered suspect due to the broad
wings of the XRT point spread function).  Assuming a thermal
Brehmsstrahlung spectrum with $kT=5.0$ keV \citep[e.g.][]{Bloom06} at
$z=0.457$, any cluster must be fainter than $2\times 10^{43}$ erg
s$^{-1}$ ($3\sigma$).  This is an order of magnitude fainter than the
rich X-ray cluster associated with GRB~050509B.

We determined the redshift of the galaxies around the host in order to
determine a possible group or cluster as found for two other SHB,
\citep{Berger07}. For most of the galaxies in the slit, we could not
determine any redshift due to a lack of strong emission or absorption
lines. We did not attempt to determine redshifts via SED modelling of
the spectral shape for these sources, as such a method would not have yielded the necessary redshift accuracy
to determine whether the galaxies could belong to a cluster or a group connected to the host galaxy.

We find that the galaxy about 5 arcsec North-West of the host,
called G3 in Fig. \ref{hostpic} and Table \ref{hostfield}, has the same
redshift as the host. Also the galaxies G2 and G5, which are about
10$-$20 arcsec from the host both have a similar redshift of z$=$0.473
and 0.471. However, this would correspond to a distance of 57$-$115 kpc from the host to G2+G5, which makes it unlikely that the host+G3
are gravitationally bound to G2+G5. G4 only shows one emission line
which, if associated with O~{\sc ii}, would give a redshift of
z$=$0.860. Together with the X-ray observations, we therefore conclude
that the host may be a member of a smaller group, but does not belong to
a larger cluster.

\subsection{Properties of Surrounding Galaxies}
A small number of the galaxies probed by the different slits show prominent
emission or absorption lines for the remaining galaxies.  The only galaxy which was identified to have the same redshift as the
host galaxy, G3, has a young stellar population. Its spectral slope is
rather flat and does not show any 4000 \AA{} Balmer break.  There is also no detection of the Ca~{\sc ii} absorption lines. The H$\beta$
equivalent width suggests an age between 4 and 20 Myrs and only one of
the Ca~{\sc ii} lines could be clearly identified. G2, which has a
slightly higher redshift than the host, shares some of its properties,
with a moderate H$\beta$ EW indicating an age between 10 and 100 Myrs
and a clear 4000 \AA{} break. However, only one of the Ca~{\sc ii}
absorption lines could be clearly identified. G4 has only one emission
line likely identified with the O~{\sc ii} doublet, which indicates a
younger stellar population. G5 has no emission lines, but clear
Ca~{\sc ii} absorption and a Balmer break, which suggests an older
galaxy.

We utilize the same methods to determine the SFR, extinction and
metallicity as those described in the previous section to determine
the properties of the other galaxies probed by the different 
slits. For G2, we get an extinction from the Balmer line ratio between
H$\beta$ and H$\gamma$ of E(B$-$V)$_{G2}$=0.96$\pm$0.7 mag. From the
flux of [O~{\sc ii}] $\lambda$ 3729, we obtain a SFR$_{G2}$ of
1.0$\pm$0.1 M$_\odot$~yr$^{-1}$ or a SSFR of 3.38
 M$_\odot$yr$^{-1}$(L/$L$*)$^{-1}$, the extinction corrected fluxes give a SFR$_{G2}$ of
61.4$\pm$3.8 M$_\odot$~yr$^{-1}$ or a SSFR of 207  M$_\odot$yr$^{-1}$(L/$L$*)$^{-1}$. The metallicity, assuming the upper branch
solution, expected for an older galaxy, is high and very similar to that of the host galaxy with 12+log(O/H)$_{G2}=$9.05 \citep{KD02} or 8.9 using the extinction corrected fluxes. For G3, we find that the extinction is
consistent with zero, which agrees with the indications of a young,
star-forming galaxy, although dusty star-burst galaxies have been
observed as well. The G3 SFR from [O~{\sc ii}] is 3.4$\pm$0.1
M$_\odot$~yr$^{-1}$ with a SSFR of approximately 21.2
 M$_\odot$yr$^{-1}$(L/$L$*)$^{-1}$ and the metallicity of either 12+log(O/H)$_{G3}=$8.23 for the lower
branch solution or 12+log(O/H)$_{G3}=$8.61 for the upper branch, which in both cases would
imply subsolar metallicities. For a clear determination of the
metallicity, the detection of H$\alpha$ or the temperature sensitive
line [O~{\sc iii}] $\lambda$ 4363 would be necessary
\citep{Izotov06}. In case the emission line identified in G4 is
      [O~{\sc ii}] $\lambda$ 3727, 3729, we get a SFR$_{G4}$ of
      4.7$\pm$0.1 M$_\odot$~yr$^{-1}$ or a SSFR of 47.0
      M$_\odot$yr$^{-1}$(L/$L$*)$^{-1}$.


\begin{table*}
 \begin{minipage}{110mm}
\caption{Emission line measurements for the host and the neighbouring galaxy}
 \begin{tabular}{@{}llllll}
  \hline 
Line & $\lambda_{obs}$ & z & EW & Flux (measured) & Flux (corrected) \\
 &  [\AA\/]  &  & [\AA\/]  & [10$^{-17}$erg/cm$^2$/s] & [10$^{-17}$erg/cm$^2$/s] \\
\hline 
{\bf Host galaxy:}&&&\\ 
$[$O~{\sc ii}$]$ $\lambda\lambda$3727,29&5429.85&0.4562&---&7.75 $\pm$ 0.22  & 1198 $\pm$ 34\\
Ca H $\lambda$ 3934&5729.02&0.4564&3.89  $\pm$  0.48&---&---\\
Ca K $\lambda$ 3969&5782.62&0.4571&8.98  $\pm$  0.58&---&---\\
H$\gamma$ 4340 & 6321.95 & 0.4565 &---&4.22 $\pm$ 0.39 & 388 $\pm$ 36\\
H$\beta$ 4861 & 7079.83 & 0.4564 & 16.94 $\pm$ 0.78 & 15.40  $\pm$  0.10 & 851 $\pm$ 6\\
$[$O~{\sc iii}$]$ 4959&7221.33& 0.4562 &---&4.86 $\pm$ 0.62 & 244 $\pm$ 31\\
$[$O~{\sc iii}$]$ 5007&7291.92& 0.4564 &---&14.1 $\pm$ 0.6 & 674 $\pm$ 29\\[2mm] \hline \\[1mm]
{\bf Host, slit2:}&&&\\ 
$[$O~{\sc ii}$]$ $\lambda\lambda$3727,29&5433.84&0.4568&---&24.9 $\pm$ 0.15 & ---\\
Ca H $\lambda$ 3934&5732.19&0.4568&3.03 $\pm$ 0.19 &---&---\\
Ca K $\lambda$ 3969&5787.39&0.4579&8.82 $\pm$ 0.21&---&---\\
H$\beta$ 4861&7084.36& 0.4569 &11.09 $\pm$ 0.95&7.29 $\pm$ 0.62 & ---\\
$[$O~{\sc iii}$]$ 4959&7225.96& 0.4567 &---&4.03 $\pm$ 0.41 & ---\\
$[$O~{\sc iii}$]$ 5007&7293.97& 0.4564 &---&10.2 $\pm$ 1.2 & ---\\[2mm] \hline \\[1mm]
{\bf G2:}&&&\\
$[$O~{\sc ii}$]$ $\lambda\lambda$3727,29&5493.76& 0.4733 &---& 8.57 $\pm$ 0.53 & 526 $\pm$ 33\\
Ca K $\lambda$ 3969&5849.10&0.4738&8.89 $\pm$ 2.32 & ---\\
H$\gamma$ 4340 & 6396.06 & 0.4735 &---&2.73 $\pm$ 0.24 & 110 $\pm$ 9\\
H$\beta$ 4861 & 7163.21 & 0.4735 & 15.92 $\pm$ 0.66& 8.93  $\pm$  0.37 & 236 $\pm$ 10\\
$[$O~{\sc iii}$]$ 4959&7307.87& 0.4736&---&2.98 $\pm$ 0.36 & 73 $\pm$ 9\\
$[$O~{\sc iii}$]$ 5007&7378.33& 0.4736 &---& 10.4 $\pm$ 0.36 & 244 $\pm$ 9\\[2mm] \hline \\[1mm]
{\bf G3:}&&&\\
$[$O~{\sc ii}$]$ $\lambda\lambda$3727,29& 5432.39&---& 0.4568 & 31.3 $\pm$ 1.5 & ---\\
H$\gamma$ 4340 &6322.35  & 0.4566 &---&6.69 $\pm$ 0.31 & ---\\
H$\beta$ 4861&7081.26&0.4567 & 46.12 $\pm$ 1.02&13.6 $\pm$ 0.3 & ---\\
$[$O~{\sc iii}$]$ 4959&7223.49&0.4567 & ---& 17.7 $\pm$ 0.5 & ---\\
$[$O~{\sc iii}$]$ 5007&7293.41& 0.4567 & ---&54.8 $\pm$ 0.5 & ---\\[2mm] \hline \\[1mm]
{\bf G4:}&&&\\
$[$O~{\sc ii}$]$ $\lambda\lambda$3727,29 &6937.05& 0.8604  &---&9.25$\pm$0.07 & ---\\[2mm] \hline \\[1mm]
{\bf G5:}&&&\\
Ca H $\lambda$ 3934&5787.39 &0.4712 & 18.54$\pm$1.21& ---&---\\
Ca K $\lambda$ 3969&5837.22 &0.4709 & 15.49$\pm$1.08& ---&---\\[1mm]

\hline 
\end{tabular} 
\label{spectable}
\end{minipage}
\end{table*}


\section{Constraints on Excess Optical Emission} 

\subsection{Constraints on an Associated Core-Collapse Supernova }
\begin{figure}
\includegraphics[width=\columnwidth, angle=0]{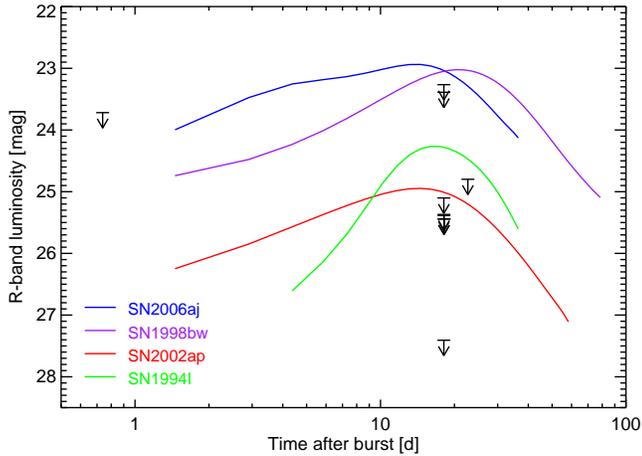}
\caption{Different SN Ic and GRB-SN templates as they would appear
  shifted to $z=0.456$, the redshift of the host galaxy. The
  SN lightcurves are time dilated accordingly, the magnitudes are
  corrected for the luminosity distance after applying a k-correction by fitting the SEDs at different
  times.  The templates are not corrected for possible extinction in the SN host galaxy. 
\label{SNlimits}}
\end{figure}

Through the use of our non-detection of the optical afterglow, we can
derive limits on the optical contribution from SN of Type Ib or Ic associated with this event.  In Figure \ref{SNlimits}, we plot three
different SN Ic templates together with the upper limits in the
$R$-band. In order to convert our photometry to rest-frame values (k-correction), we fit the SED (in flux units) derived from data
points and interpolations of the $UBVRI$ lightcurve for each day,
redshifting the SED to $z=0.456$, and determined the new fluxes at the
effective $R$-band wavelength before transforming these flux values back into
magnitudes.  Furthermore, the templates were corrected for
time dilation as well as for Galactic extinction.
The comparison templates used are the ``standard'' long-duration
GRB-SN 1998bw \citep{Galama98}, an SN associated with an X-ray flash (XRF), SN 2006aj \citep{Sollerman06},
and faint Ic supernovae 2002ap \citep{Foley03} and 1994i \citep{Richmond96}.  All observed SNe that
have been associated with a long-duration GRB are within 0.5 mag of the
peak luminosity of SN 1998bw, although some XRF-SNe have appeared to
be significantly fainter, possibly because of dust (Soderberg et
al. 2006, see also Woosley \& Bloom 2006).

Taking the deep limit in the R-band from Keck I on Aug 11, 2007, of
$R=27.4$ mag (2$\sigma$), any broad-line SN Ic associated with this
short GRB would have been at least 100 times fainter than SN 1998bw
and still 10 times fainter than SN Ic 1997ef.  If the extinction in
the host galaxy is indeed as high as derived in
Sect$.$ \ref{extinction}, a very faint SN Ic could indeed have been
missed, SN 1998bw would, however, still have been brighter than the
observed limits.

\subsection{Li-Paczy$\mathbf{\acute{n}}$ski Modeling}

Our observations also allow us to place limits on the optical emission
from a so-called ``mini-SN" or LP-SN \citep{LiPaczynski98, Rosswog02, Kulkarni05}.
In this model, the coalescence of two compact objects produces a sub-relativistic ejecta of nuclear dense material comprising
approximately $10^{-4}-10^{-2} M_{\sun}$, depending on the nuclear
equation of state and the properties of the merger (such as nature of
the components and initial mass ratio).  The rapid decompression of
the ejecta as it expands adiabatically is thought to result in the
synthesis of a variety of radioactive elements, which could decay over
a wide range of timescales.  With nominal and simplistic assumptions,
the spectrum of such emission is expected to be quasi-thermal and peak
in the optical/UV range with a characteristic timescale of about 1
day.

\begin{figure}
\includegraphics[width=\columnwidth, angle=0]{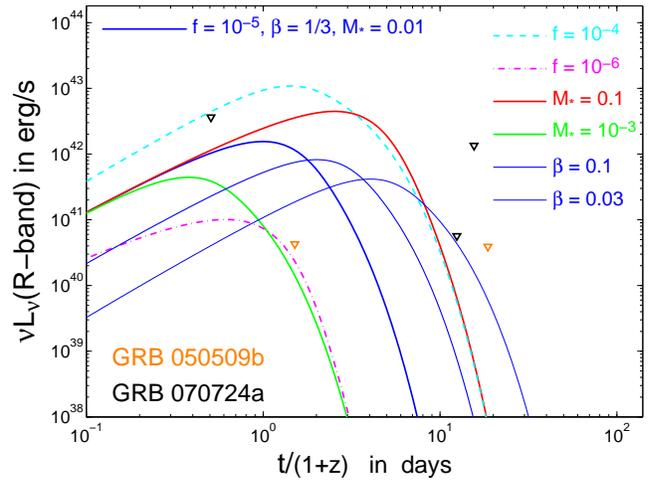}
\caption{Li \& Paczynski models for different ejecta masses M$_*$,
  velocities $\beta$, and energy conversion factor $f$.  Our $R$-band
  upper limits are shown in black along with the upper limits
  presented by \citet{Hjorth05b} for GRB~050509b.
\label{mini-SNlimits}}
\end{figure}

Using the simple derivations presented in \citet{LiPaczynski98}, we
can construct an analytic model to describe the properties of this
LP-SN light curve as a function of three parameters: the mass, $M$,
and velocity, $\beta c$, of the ejecta, and the fraction of the ejecta
energy, $f$, that goes into radioactive decay per $e$-fold in time over the relevant timescales.  Following
\citet{LiPaczynski98}, the peak luminosity, time to reach this value,
and effective temperature of the ejecta, can be given respectively as:
\begin{eqnarray} \label{eq:Lum}
L_{\rm pk} =
2.1~\times~10^{44}~\rm{ergs~s^{-1}}~~~~~~~~~~~~~~~~~~~~~~~~~~~~~~~~~~
& \nonumber
\\ \times~\left(\frac{f}{0.001}\right)\left(\frac{M}{0.01M_{\sun}}\right)^{1/2}\left(3\beta\right)^{1/2}\left(\frac{\kappa}{\kappa_{e}}\right)^{-1/2},
\end{eqnarray}
\begin{equation} \label{eq:tpk}
t_{\rm pk} = 0.98~{\rm days}
\left(\frac{M}{0.01M_{\sun}}\right)^{1/2}\left(3\beta\right)^{-1/2}\left(\frac{\kappa}{\kappa_{e}}\right)^{1/2},
\end{equation}
and
\begin{eqnarray} \label{eq:Temp}
T_{\rm eff,pk} =
2.5~\times~10^{4}~\rm{K}~~~~~~~~~~~~~~~~~~~~~~~~~~~~~~~~~~~~~~~~~~~~~~
& \nonumber
\\ ~~~~~\times~\left(\frac{f}{0.001}\right)^{1/4}\left(\frac{M}{0.01M_{\sun}}\right)^{-1/8}\left(3\beta\right)^{-1/8}\left(\frac{\kappa}{\kappa_{e}}\right)^{-3/8}.
\end{eqnarray}

Here $\kappa$ is the average opacity ($\kappa_e$ is the opacity
caused by electron scattering).   The time of peak luminosity occurs when the photons can diffuse outward on the dynamical time, so that most of
Êthe thermal energy that is produced by the radioactive decay can be
Êradiated efficiently before suffering significant adiabatic cooling.
ÊSince the photon diffusion time is $\sim \tau$ times larger than the
Êsource light crossing time, $R/c$, and $R = \beta c t$, we can
Êequate the dynamical time $t_{\rm dyn} \sim t = R/\beta c$ with $\tau
ÊR/c$ at $t_{\rm pk}$ and therefore $\beta^{-1} \sim \tau(t_{\rm pk}) \propto
Ê\kappa M/(\beta t_{\rm pk})^2$ and therefore $t_{\rm pk} \propto (\kappa
ÊM/\beta)^{1/2}$. The peak luminosity is approximately given by
Ê$L_{\rm pk} \sim f M c^2/t_{\rm pk} \propto f(\beta M/\kappa)^{1/2}$, and
Êthe effective temperature is obtained by equating this luminosity to
Êthat of a black body, $L_{\rm pk} \sim 4\pi R^2(t_{\rm pk})\sigma
ÊT_{\rm eff,pk}^4 = 4\pi (\beta c t_{\rm pk})^2\sigma T_{\rm eff,pk}^4 $ and
Êtherefore $T_{\rm eff,pk} \propto L_{\rm pk}^{1/2}(\beta t_{\rm pk})^{-1/2}
Ê\propto f^{1/4}\kappa^{-3/8}(\beta M)^{-1/8}$.

As a result, the luminosity of the resulting light curve is directly proportional to the fraction of the ejecta energy
that goes into radioactive decay (i.e. the ejecta efficiency).  An
increase in the assumed mass in the ejecta has the effect of both
increasing the total luminosity of the emission as well as increasing
the time, $t_{\rm pk}$, to peak luminosity, $L_{\rm pk}$.  On the
other hand, an increase of ejecta velocity leads to an increase in
the peak luminosity but a decrease in the peak duration.

The most efficient conversion of nuclear energy to observable
luminosity is provided by elements with a decay timescale comparable
to the timescale it takes the ejected debris to become optically thin,
$t_\tau$. In reality, there is likely to be a large number of nuclides
with a very broad range of decay timescales.  The ejecta is optically
thick at early stages but its optical depth falls rapidly as the
ejecta expands.  As this happens, radiative losses at the photosphere
become important and the peak emission occurs when the optical depth
reaches $\sim$ $1/\beta$. Current observational limits thus place interesting
constraints on the abundances and the lifetimes of the radioactive
nuclides that form in the rapid decompression of nuclear-density
matter -- they should be either very short or very long when compared
to $t_\tau$ so that radioactivity is inefficient in generating a high
luminosity.

Figure \ref{mini-SNlimits} shows the calculated $R$-band light curves
for a range of ejecta mass $M$, velocity $\beta$ and the fraction of the ejecta
energy $f$ that goes into radioactive decay, effectively the energy conversion factor.  Plotted as black downward arrows are three $R$-band
upper limits taken at roughly 0.74, 18.14, and 22.70 days from left to
right.  For comparison, we have also plotted the $R$-band upper limits
for GRB~050509b presented by Hjorth et al. 2005b and Bloom et
al. 2006.  Unfortunately, many of our observations between days 2 and
18 are less constraining, so that we can only place limits on a
very fast, massive, and highly efficient ejecta.  Likewise, the deep
upper limit at late time can only constrain a very slow, high mass
ejecta.

\begin{figure}
\includegraphics[width=\columnwidth, angle=0]{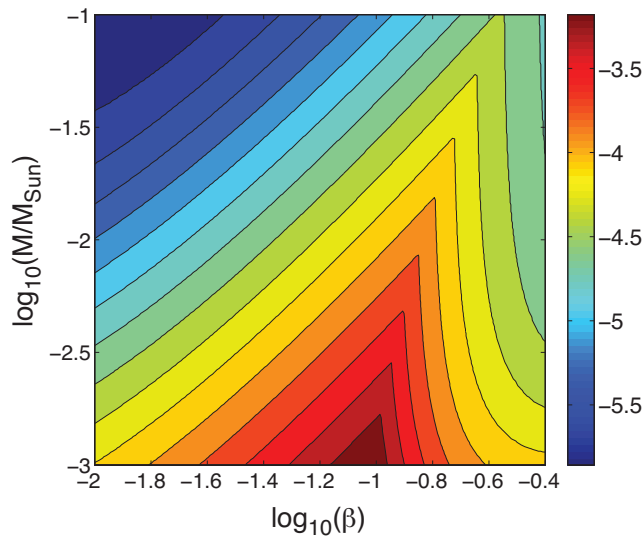}
\caption{A parameter space plot for GRB~070724A showing the
  maximal allowed value that our observations place on the energy conversion
  factor $f$ for a range of ejecta masses $M$ and velocities
  $\beta$. For a given $M_{ejecta}$ and $\beta$, the color (coded at
  right) corresponds to the maximum allowed value for $f$.
\label{Fig:pspace1}}
\end{figure}

\begin{figure}
\includegraphics[width=\columnwidth, angle=0]{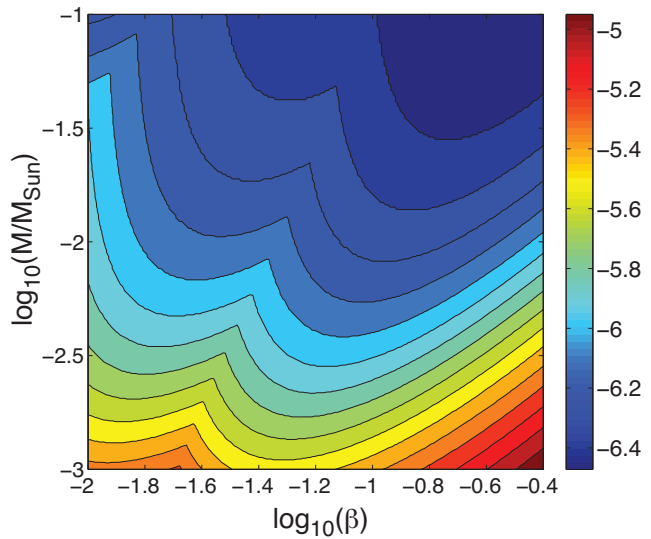}
\caption{A parameter space plot for GRB~050509b showing the
  maximal allowed value that the observations made by \citet{Hjorth05b} place on
  the energy conversion factor $f$ for a range of ejecta masses
  $M_{ejecta}$ and velocities $\beta$.  For a given $M_{ejecta}$ and
  $\beta$, the color (coded at right) corresponds to the maximum
  allowed value for $f$.
\label{Fig:pspace2}}
\end{figure}

In Figures \ref{Fig:pspace1} \& \ref{Fig:pspace2}, we show the
parameter space excluded by our observations for GRB~070724A as well
as the upper limits on emission from GRB~050509b placed by
\citet{Hjorth05b}.  The areas in parameter space which are shown are
divided into regions that are dominated by the constraint from a small
number of upper limits. The dividing lines between these regions are
demarcated by sharp changes in the direction of the contour lines.

In Figure \ref{Fig:pspace1}, the right to lower right side of the plot
is dominated by the upper limit of 23.72 mag in $R$-band at 0.5068
days, while the other half of the plot is dominated by the upper limit
of 27.4 mag in $R$-band at 12.44 days, both epochs listed in the host
frame.  The darker blue regions of the plot represent deeper limits on
the fraction of the ejecta energy $f$ that goes into radioactive
decay.  The parameter space that is least constrained represents the
intermediate velocity ejecta at a wide range of masses.  This
corresponds to the time range for peak between days 2 and 18 for which
we have only shallow upper limits in magnitude.

For comparison, the upper limits made by \citet{Hjorth05b} place
stricter limits on the presence of a LP-SN for GRB~050509b.  The
derived limits for GRB~050509b shown in Figure \ref{Fig:pspace2} are
dominated by the upper limit of 26.6 mag in the $R$-band at 1.85 days
(lower right) and by the upper limit of 27.5 mag in the $V$-band at
3.92 days (upper left). At the very upper left corner (high $M$ and
low $\beta$) there is a small region which is dominated by the upper
limit of 26.7 mag in $R$-band at 22.83 days.  This is natural since $t_{pk} \propto (M/\beta)^{1/2}$ (see eq. 2).

\section{Discussion}


We present the results of a deep and extensive observing campaign of the short
Êduration hard spectrum (SGRB) GRB~070724A.  Although our observations do not reveal an
optical or NIR transient associated with this event, the derived upper
limits can constrain optical emission from a traditional forward
shock as well as the physical parameters of SNe, or SNe-like models.
Our early NOT and late time Keck observations show no sign of the
optical emission from a forward shock in the slow cooling
regime, as predicted by the fireball model.  However, this optical
contribution could have easily eluded detection given the weakness of the event combined with the potential extinction by the host galaxy.  


\begin{figure*}
\includegraphics[width=7in,angle=0]{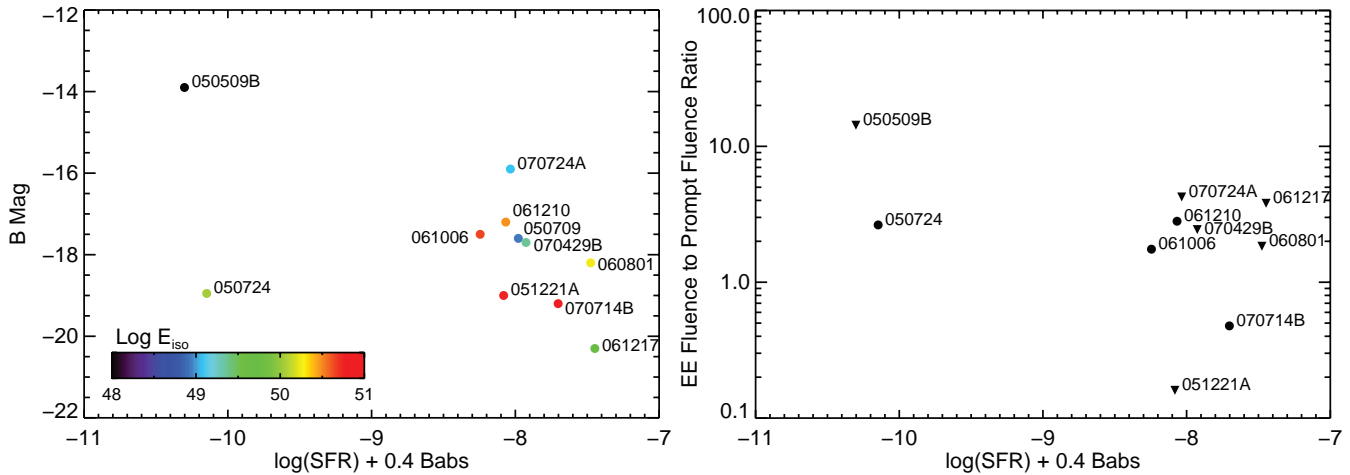}
\caption{Left: The R$_{\rm c}$ magnitude of the afterglow (or upper limit thereto) at 1 day after the GRB in a common $z = 1$ frame adopted from \citet{Kann08}. vs$.$ the specific star formation rate of the host galaxy as presented by \citet{Rhoads08}.   GRB~070724A represents the deepest limit on the existence of short time scale supernova-like emission for a short GRB in a moderately star forming galaxy. The color scheme represents the isotropic equivalent energy of the event. Right: The ratio of the extended emission fluence to that of the initial spike (or limit thereof)
vs$.$ the host's specific star formation rate of a subset of short GRBs with known redshift.  There is no clear connection between the events with
significant extended emission and the properties of the host galaxy.
\label{Fig:Eiso_SSFR_Afterglow}}
\end{figure*}

Given the measured reddening value due to the host of E(B$-$V)$=$1.2$\pm$0.5 mag, any optical emission could have suffered as much as $A_{V} \sim 4$ magnitudes of extinction if the burst occurred well within the galaxy's core.  In this case, the optical limits on a mini-SN quickly become unconstrained and our late time observations would likewise barely constrain even most broad-lined Type Ic SNe.  If the GRB exploded at a large offset to the host galaxy, as has been seen in other short GRB events, or if it the event occurred on the near side rather than the far side of the host, than much of the measured extinction would be largely irrelevant.  The measured $N_{H}$ could potentially help to resolve these different scenarios.  Of the two Swift events which have solidly been shown to have $A_{V} > 3$, both have had large $N_{H}$ column density at late times, of the order $5 \times 10^{22}$ \citep{Perley09}.  We measure a similar column densities at $t < 300$ sec in association with large hardness variations in the X-ray detections, suggesting that the excess column density does not reflect the intrinsic absorption of the host.  At late times ($t > 300$ sec), the $N_{H}$ value becomes consistent with the Galactic value $N_{\rm H, Galactic}= 1.2 \times
10^{20}$ cm$^{-2}$ \citep{Dickey90}.  Furthermore, the extinction value is derived from emission line measurement, indicating the extinction towards the regions of the galaxy that harbor younger populations of stars.  These regions may differ greatly in their dust content from the extended disk or halo in which one would expect to find the older population of stars which are thought to be the progenitors of SHBs.  Therefore we conclude that the site of GRB~070724a was likely not affected by large amounts of extinction, although this remains a caveat in the proceeding discussion. 

Because our limiting magnitudes are deepest at early and late times,
our observations primarily constrain an efficient, very slow or a very
fast ejecta with mass higher than $0.01M_{\sun}$.  Our observations do
little to exclude a low mass ejecta at intermediate velocities
resulting in emission reaching peak luminosity near 1 day.  This can
be seen in the case of GRB~050509b, where two limiting magnitudes in
$R$-band are of similar depth although the earlier observation is far
more constraining.  This is due to the characteristic timescales
associated with the peak emission predicted by the LP-SN model.  This
underscores the importance of deep optical observations hours to days
after the short-hard GRB. To be sure, our calculations are based
on a very simplified analytic model, so it is not clear whether this
basic result would hold for more realistic models, e.g. \citet{Kulkarni05}.

We note that GRB~050509b and GRB~070724A represent the fainter end of
the energetics distribution \citep{Kann08}, at $2.40 \times 10^{48}$
and $1.54 \times 10^{49}$ ergs respectively, for short bursts thus far
detected by Swift.  If there is a correlation between the energy
released in the prompt gamma-ray emission and the velocity and/or mass
of the resulting ejecta, not an unreasonable scenario, then one would
expect more energetic bursts to result in brighter emission from a
mini-SN.  Therefore, deep optical observations of more energetic SHB
events may yet yield detectable emission.

To date, only 14 short bursts (GRB~050709,
GRB~050724, GRB~051221A, GRB~060121, GRB~060313,
GRB~061006, GRB~061201, GRB~070707, GRB~070714B, GRB~070809, GRB~071227, GRB~080503, GRB~080905A, GRB~090510 ) have resulted in the
detection of an optical afterglow and therefore a position determined
at subarcsecond level. Only four short GRBs detected thus far
were proposed to be associated with early type galaxies such as elliptical
or early type spiral galaxies, namely GRB~050509B \citep{Gehrels05,
  Bloom06}, GRB~050724 \citep{Barthelmy05, Berger05, Gorosabel06,
  Malesani07}, and possibly GRB~050813 \citep{Prochaska06, Ferrero07,
  Berger06} and GRB~060502B \citep{Bloom06}. The lack of emission
lines in these galaxies implies low limits on any ongoing
star-formation, the age of their stellar populations is of the order
of Gyrs, and their metallicity is around solar. All other short GRB
hosts are late type or irregular, star forming galaxies, although with
lower star formation rates as compared to their long GRB host
counterparts. The best studied examples of late type galaxies associated with short bursts to date are the irregular host
of GRB~050709 \citep{Hjorth05b, Covino06, Fox05} and the host of
GRB~051221A \citep{Soderberg06}.

The host galaxy of GRB~070724a is one of the growing number of star forming
galaxies to harbor a short GRB. Despite the lack of an optical
detection, the position of the XRT error circle covering most of the
galaxy makes an association very likely.  The host galaxies of short
bursts have been found to be a rather diverse population, in contrast
to the more uniform sample of star forming galaxies or at least
star-forming regions within larger galaxies for long-duration GRB
hosts \citep{Christensen04, Fruchter06}. Recently, \cite{Berger09}
published properties for the full set of short GRB host galaxies,
whose properties could be determined. They find that the range of SFRs is
moderate with 1--10 M$_\odot$~yr$^{-1}$, with some outliers with very low
SFR, which is slightly lower than found for long duration GRBs
\citep[e.g.][]{Christensen04}. The metallicity is in general 0.6 dex \citep{Modjaz08}
higher than for long-duration GRB hosts, however, this is based
on a rather small sample for nearby long duration GRBs. The
host of 070724A has quite average properties compared to the short GRB sample presented in \cite{Berger09} in terms of SFR and metallicity.

This diversity in both the host galaxy population and also the prompt
energetics of short GRBs hints at the possibility of progenitors that
may not be entirely restricted to the most widely favored scenario
involving the merger of compact binaries.  The dichotomy between the
specific star formation rates of short GRBs hosts, in particular, has
been proposed by \citet{Rhoads08} as a possible indication of multiple
progenitors.  Unlike the optical limits previously obtained for
GRB~050509b \citep{Hjorth05b}, which occurred within a quiescent galaxy
with little or no ongoing star formation, our observations of
GRB~070724A, along with those of GRB~050709, place the deepest
constraints on the presence of short timescale supernova-like emission of a short GRB
in a moderately star forming host galaxy.

This can be seen in the left panel of Figures \ref{Fig:Eiso_SSFR_Afterglow}, where we plot the R$_{\rm c}$ magnitude of
the afterglow (or corresponding upper limit) 1 day after the GRB in a common $z = 1$ frame, adopted from \citet{Kann08}. vs$.$ the specific star formation rate of the host galaxy for a sample of short GRBs with known redshift estimated by
\citet{Rhoads08}.  The color scheme in Figure
\ref{Fig:Eiso_SSFR_Afterglow} represents the isotropic equivelent energy $E_{\rm iso}$ of the event  This effectively
sets a limit to short timescale emission from a mini-SN, as any
emission in excess to the afterglow light curve would have been
detected.  The constraints on short timescale supernova like emission are deepest for GRB~050509B and GRB~070724A
for quiescent and moderately star forming galaxies respectively.

The extended emission observed to follow the initial gamma-ray spike
in many events detected by Swift and BATSE has also been suggested as
a possible indicator of progenitor diversity within the short GRB
population.  This phenomenon can generally be characterized by
additional gamma-ray emission lasting several tens of seconds and
fluence values that in some cases exceed that of the initial short GRB
spike.  The nature of this emission has recently been discussed in
detail by \citet{Perley08} who found that roughly 30$\%$ of short GRBs
detected by Swift exhibit it in some form.  They also find that the
extended-to-prompt fluence ratio exhibits a large variance, with no
clear correlation between the brightness or fluence of the initial
spike and that of the subsequent extended emission.

We plot the ratio of the extended
emission fluence to that of the initial spike (or limit thereof)
adopted from \citet{Perley08} vs$.$ the specific star formation rate
of a subset of short GRBs with known redshift, including
GRB~070724A, in the right pane of Figure \ref{Fig:Eiso_SSFR_Afterglow} . There is no clear correlation between the events with
significant extended emission and the type of galaxy harboring them,
although the limited sample size makes the analysis far from
conclusive.  Despite this, placing such constraints on additional
optical and gamma-ray components of short GRBs in galaxies of varying
star formation rates and hence stellar populations begins to give us
insights into the possible diversity of their progenitors.

Unless short GRBs are eventually found to be accompanied by tell-tail
emission features like the supernovae associated with long-duration
GRBs, the only definitive understanding of their progenitors will come
from possible associations to direct gravitational or neutrino
signals.  Therefore, continued attempts to observe optical components
from processes not associated with the standard afterglow emission
from short GRB remains vital to our understanding of the nature of
these events.

\section*{Acknowledgments}

D.K. acknowledges financial supported through the NSF Astronomy $\&$
Astrophysics Postdoctoral Fellowships under award
AST-0502502 and the Fermi Guest Investigator program. C.T. wants to thank Joshua Bloom and the Berkeley
GRB group for their hospitality while most of this work was done.  N.B. gratefully acknowledges support from a Townes
Fellowship at U. C. Berkeley Space Sciences Laboratory and partial
support from J. Bloom and A. Filippenko.  J. G. gratefully acknowledges a
ÊRoyal Society Wolfson Research Merit Award.  Based on observations made
with the Nordic Optical Telescope, operated on the island of La Palma
jointly by Denmark, Finland, Iceland, Norway, and Sweden and with the
William Herschel Telescope in the Spanish Observatorio del Roque de
los Muchachos of the Instituto de Astrof\'{\i}sica de Canarias. We
thank the VLT staff for excellent support performing the ToO
observations of the host galaxy.  Collection of SMARTS data is supported by NSF-AST 0707627.  This work was supported in part by the U.S. Department of Energy contract to SLAC no. DE-AC3-76SF00515.

\bibliographystyle{apj}
\bibliography{GRB070724ver11}
\label{lastpage}

\end{document}